\documentclass[twocolumn,english,amsmath,amssymb,aps,showpacs,prl]{revtex4-1}
\usepackage[latin9]{inputenc}
\usepackage{babel}
\usepackage{amsmath}
\usepackage{amssymb}
\usepackage{graphicx}
\usepackage{esint}
\usepackage[unicode=true,pdfusetitle,
 bookmarks=true,bookmarksnumbered=false,bookmarksopen=false,
 breaklinks=false,pdfborder={0 0 1},backref=false,colorlinks=false]
 {hyperref}
\hypersetup{
 colorlinks,linkcolor=blue,citecolor=blue,urlcolor=blue}

\makeatletter
\usepackage{babel}
\usepackage{babel}
\usepackage{mathrsfs}

\makeatother

\begin{document}

\title{Numerical calculation of the Casimir\textendash Polder interaction
between a graphene sheet with vacancies and an atom }

\author{T. P. Cysne}

\affiliation{Instituto de F\'\i sica, Universidade Federal do Rio de Janeiro, Caixa
Postal 68528, Rio de Janeiro 21941-972, RJ, Brazil}

\author{T. G. Rappoport}

\affiliation{Instituto de F\'\i sica, Universidade Federal do Rio de Janeiro, Caixa
Postal 68528, Rio de Janeiro 21941-972, RJ, Brazil}

\author{Aires Ferreira}

\affiliation{Department of Physics, University of York, York YO10 5DD, United
Kingdom}

\author{J. M. Viana Parente
 Lopes}

\affiliation{Centro de Fisica and Departamento de Fisica, Universidade do Minho,
Campus de Gualtar, Braga 4710-057, Portugal}
\affiliation{Departamento de F\'\i sica e Astronomia,Faculdade de Ci\^encias da Universidade do Porto,
Rua do Campo Alegre 687, 4169-007 Porto, Portugal}

\author{N. M. R. Peres}

\affiliation{Centro de Fisica and Departamento de Fisica, Universidade do Minho,
Campus de Gualtar, Braga 4710-057, Portugal}
\begin{abstract}
In this work the Casimir\textendash Polder interaction energy between
a rubidium atom and a disordered graphene sheet is investigated \textit{\emph{beyond
the Dirac cone approximation}} by means of accurate real-space tight-binding
calculations. As a model of defected graphene, we consider a tight-binding
model of $\pi$-electrons on a honeycomb lattice with a small concentration
of vacancies. The optical response of the graphene sheet is evaluated
with full spectral resolution by means of exact Chebyshev polynomial
expansions of the Kubo formula in large lattices with in excess of
ten million atoms. At low temperatures, the optical response of defected
graphene is found to display two qualitatively distinct behavior with
a clear transition around finite (non-zero) Fermi energy. In the vicinity
of the Dirac point, the imaginary part of optical conductivity is
negative for low frequencies while the real part is strongly suppressed.
On the other hand, for high doping, it has the same features found
in the Drude model within the Dirac cone approximation, namely, a
Drude peak at small frequencies and a change of sign in the imaginary
part above the interband threshold. These characteristics translate
into a non-monotonic behavior of the Casimir\textendash Polder interaction
energy with very small variation with doping in the vicinity of the
neutrality point while having the same form of the interaction calculated
with Drude's model at high electronic density. 
\end{abstract}
\maketitle

\section{Introduction}

\vspace{0.1in}

Dispersive forces\textemdash including van der Waals, Casimir, and
Casimir\textendash Polder types\textemdash are interactions between
neutral, but polarizable objects, and have their origin in fluctuations
of the vacuum electromagnetic field~\cite{reviews}. The Casimir
force involves interactions between macroscopic objects~\cite{casimir},
such as plates, while Casimir\textendash Polder forces act between
a macroscopic object and a microscopic particle~\cite{casimirpolder}.
Van der Waals forces act between objects in the short range regime,
where effects of retardation can be neglected. These forces are dominant
on nano and micro scales and their control and manipulation are important
to applications, such as nano-electromechanical systems, among others~\cite{nem,nem2}.
Dispersive forces are strongly influenced by the shape and material
composition, as well as the dielectric and magnetic responses of the
objects they act upon. It is possible to tailor the sign and magnitude
of dispersive forces by tuning, for example, the dielectric response
of the plate. As a result, the correct modeling of dispersive forces
from a materials science perspective becomes important~\cite{LiliaWoods-Review-2015,Galina-Review_2009}.

Since its isolation, graphene has attracted great attention, owing
to its unconventional low-energy physics described by the Dirac\textendash Weyl
equation for massless excitations in two spatial dimensions, and a
number of desirable physical properties, including superior mechanical
strength, high charge carrier mobilities, and gate-tunable optical
response~\cite{rgraphene1,rgraphene2,rgraphene3}. Intense theoretical
effort has been devoted to the study of Casimir \cite{Bordag2009,Bordag2012,Drosdoff,Fialkovsky,Galina2013,Gomez,MacdonaldPRL,Sernelius,Svetovoy,Klimchitskaya-Mostepanenko-14,Galina2014,Bordag2016,Drosdoff2012-2}
and Casimir\textendash Polder \cite{Churkin,Galina2014,Chaichia3,Judd,Sofia-Scheel-13,Cysne_2014,LiliaWoods-2016}
interactions in graphene and related systems~\cite{CasimirChern,abinitio}.
The Casimir\textendash Polder energy of different atoms on single
layer has been considered in Refs.~\cite{Churkin,Chaichia3}, and
on multilayer graphene in Ref.~\cite{LiliaWoods-2016}. The tunability
of interactions have been demonstrated in atom-on-graphene \cite{Cysne_2014}
and in graphene bilayers \cite{MacdonaldPRL} using external magnetic
fields, and in a graphene\textendash metal system by tuning the chemical
potential \cite{Bordag2016}. In general, a Dirac cone approximation
is considered where the reflection coefficients of graphene are calculated
either within the hydrodynamic model or the polarization tensor with
a Drude model approach. The weakness of dispersive interactions on
graphene systems is experimentally challenging, and most of theoretical
predictions point to an enhancement of interactions by charge doping
\cite{Drosdoff2012-2,Bordag2016}. Recently, the control of the interaction
between graphene and naphtalene molecule at short distances (van der
Waals regime) has been achieved exploiting the high tunability of
the chemical potential \cite{Huttman-2015}.

Recent advances in the understanding of dispersive interactions involving
graphene and other low-dimensional systems have shown the importance
of a detailed characterization the electrical response of the layers
for the control and tailoring of dispersive forces. Although \emph{ab
initio} methods have been used to model van der Waals forces~\cite{abinitio},
a more materials oriented approach to study Casimir interactions is
still needed. In this article, we consider a realistic model of a
large graphene sheet with vacancies. To determine the Fermi energy
dependence of its optical conductivity, we employ an accurate large-scale
quantum transport approach based on an exact polynomial representation
of disordered Green functions recently introduced in Ref.~\cite{Ferreira15}.
The large number of expansion moments in the numerical evaluation
of the Kubo formula in large graphene lattices allows us to determine
the optical response with fine spectral resolution. This information
is then used to compute the Casimir\textendash Polder force between
a defected graphene sheet and an atom in function of the charge doping,
and compare it with the force calculated using the Drude model. Far
from the Dirac point, the Casimir\textendash Polder force varies linearly
with the chemical potential. The Drude model is found in accord with
numerical calculations in that regime, as expected, but fails to capture
the behavior of the Casimir\textendash Polder force close to the Dirac
point. Furthermore, we find that the strength of the interaction is
reduced in the vicinity of the Dirac point, following the trend of
the dc conductivity~\cite{Ferreira15}, and increases again above
a certain Fermi energy scale $\mu^{*}>0$, in contrast with the monotonic
enhancement of interactions predicted by calculations based on perfect
graphene models.

This article is organized as follows. Section~\ref{sec2:Methodology}
outlines the real space quantum transport methodology used to extract
the optical conductivity of large disordered graphene lattices. In
Sec.~\ref{sec:Optical-Conductivity}, we apply the methodology to
a graphene lattice with a dilute concentration of vacancies. In section~\ref{sec3}
we describe the calculation of the Casimir\textendash Polder force
between the graphene layer and a rubidium atom and present our results.
Finally, section~\ref{sec4} summarizes the main findings of our
work.

\section{METHODOLOGY\label{sec2:Methodology}}

\begin{figure}
\centering{}\vspace{0.1in}
 \includegraphics[width=0.8\columnwidth]{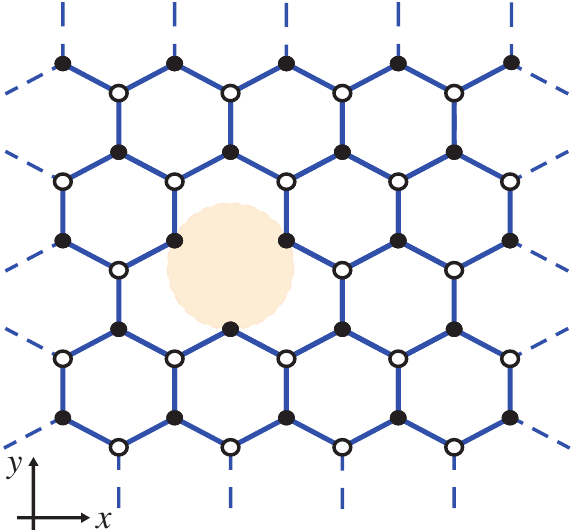}\caption{\label{fig:01_schematic}Schematic of a graphene lattice with vacancy
defects. $A$($B$) sublattices are represented by filled (open) circles.
Shaded area shows a vacancy. The numerical simulations in this work
have a computational domain of size $3200\times3200$, with periodic
boundary conditions on both directions (torus).}
\end{figure}

The graphene sheet is modeled by a tight-binding Hamiltonian of $\pi$~electrons
defined on a honeycomb lattice 
\begin{equation}
\hat{H}=-t\sum_{\langle i,j\rangle}\:(\hat{a}_{i}^{\dagger}\hat{b}_{j}+\textrm{H.c.})\,,\label{eq:TB-graphene}
\end{equation}
where the operator $a_{i}^{\dagger}$ creates an electron at site
$\mathbf{r}_{i}=(x_{i},y_{i})$ on sublattice $A$ (an equivalent
definition holds for sublattice $B$), and $t=2.7$~eV is the nearest-neighbor
hopping integral \cite{RMPgraphene}. The point defects are introduced
by removing sites in any sublattice at random (compensated vacancies).
The defect concentration is $n_{i}=N_{d}/D$, where $N_{d}$ is the
number of missing carbon atoms and $D$ is the number of sites in
the pristine lattice (see Fig.~\ref{fig:01_schematic}).

The real part of the diagonal optical conductivity at zero temperature
and finite frequency is given by \cite{Mahan} 
\begin{equation}
\Re\,\sigma(\omega)=\frac{\pi}{\omega\,\Omega}\int_{\mu-\hbar\omega}^{\mu}d\epsilon\,\,\textrm{Tr}\,\langle\hat{J}_{x}\,\hat{A}(\epsilon)\,\hat{J}_{x}\,\hat{A}(\epsilon+\hbar\omega)\rangle_{\textrm{c}}\,,\label{eq:optical_conductivity}
\end{equation}
where $\hat{J}_{x}=(ite/\hbar)\sum_{\langle i,j\rangle}(x_{i}-x_{j})(\hat{a}_{i}^{\dagger}\hat{b}_{j}-\textrm{H.c.})$
is the $x$-component of the current density operator, and 
\begin{equation}
\hat{A}(\epsilon)=-\frac{1}{\pi}\,\Im\,\frac{1}{\epsilon-\hat{H}+i\eta}\,,\label{eq:spectral operator}
\end{equation}
is the spectral operator of the system. The symbol $\langle...\rangle_{\textrm{c}}$
denotes configurational average, $\Omega$ is the area of the lattice,
$\mu$ is the chemical potential, and $\eta$ is a small broadening
parameter required for numerical convergence. Physically, the broadening
$\eta=\hbar/\tau_{i}$ mimics the effect of uncorrelated inelastic
scattering processes with lifetime $\tau_{i}$ (e.g., due to phonons),
and can be viewed as an energy uncertainty due to coupling of electrons
to a bath \cite{Thouless_81,Imry}.

The response functions of large tight-binding systems can be assessed
numerically by means of specialized spectral methods \cite{spectral_Kosloff84,spectral_Roche_Mayou_97,spectral_Tanaka98,spectral_Weisse04,spectral_Yuan10,spectral_Holzner11}.
A particularly convenient approach is the kernel polynomial method
\cite{KPM}, in which spectral operators are approximated by accurate
matrix polynomial expansions. The coefficients of the polynomial expansion
are computed recursively thereby bypassing matrix inversion that limits
systems sizes in exact diagonalization schemes. The kernel polynomial
method has been applied intensively to study the electronic properties
of disordered graphene \cite{KPM_G_Ferreira11,KPM_G_Fan14,KPM_G_Garcia15,ReviewChebG}.
Here, we make use of an exact Chebyshev polynomial representation
of the resolvent operator recently obtained in Ref.~\cite{Ferreira15},
in order to perform numerically exact large-scale calculations of
the optical conductivity. The starting point in our approach is the
operator identity $(z-\hat{h})^{-1}=\sum_{n=0}^{\infty}a_{n}(z)\,\mathcal{T}_{n}(\hat{h})\,,$
where $z=(\epsilon+i\eta)/W$, $\hat{h}$ is the rescaled Hamiltonian
of disordered graphene $\hat{h}=\hat{H}/W$ (here $W=3t$ is half
bandwidth), and $\mathcal{T}_{n}(\hat{h})$ are matrix Chebyshev polynomials
of first kind (see Appendix). Using this expansion, the spectral operator
{[}Eq.~(\ref{eq:spectral operator}){]} can be recast into the form
\begin{equation}
\hat{A}=-\frac{1}{\pi W}\sum_{n=0}^{\infty}\Im[a_{n}(z)]\,\mathcal{T}_{n}(\hat{h})\,,\label{eq:spectral_op_expansion}
\end{equation}
whose action on a given basis set can be computed iteratively by standard
Chebyshev recursion \cite{KPM}.

In a numerical implementation, the sum in Eq.~(\ref{eq:spectral_op_expansion})
is truncated when convergence to a given desired accuracy is achieved.
The $N$th-order approximation to the optical conductivity is therefore
given by

\begin{equation}
\Re\:\sigma^{(N)}(\omega)=\frac{\pi}{\,\omega\,\Omega}\sum_{n,m=0}^{N-1}\sigma_{nm}\,A_{nm}(\mu,\omega)\,,\label{eq:Re_sigma_Chebyshev}
\end{equation}
where 
\begin{align}
\sigma_{nm} & =\textrm{Tr}\:\langle\hat{J}_{x}\,\mathcal{T}_{n}(\hat{h})\,\hat{J}_{x}\,\mathcal{T}_{m}(\hat{h})\rangle_{\textrm{c}}\,,\label{eq:Gamma_nm}\\
A_{nm}(\mu,\omega) & =\frac{1}{\pi^{2}W^{2}}\int_{\mu-\hbar\omega}^{\mu}d\epsilon\,\alpha_{n}(\epsilon)\alpha_{m}(\epsilon+\hbar\omega)\,,\label{eq:Anm}
\end{align}
and $\alpha_{n}(\epsilon)$ is a shorthand for $\Im[a_{n}\left((\epsilon+i\eta)/W\right)]$.
Clearly, the problem boils down to the evaluation of $\sigma_{nm}$,
which contains the relevant dynamical information. Once the expansion
moments have been determined, the optical conductivity can be quickly
retrieved using Eq.~(\ref{eq:Re_sigma_Chebyshev}). For a recent
review on the application of Chebyshev expansions in the context of
disordered graphene, we refer the reader to Ref.~\cite{ReviewChebG}.

\subsection{Optical Conductivity of Disordered Graphene\label{sec:Optical-Conductivity}}

With the approach described in the previous section we can study,
in a numerically rigorous way, the optical conductivity of graphene
in the presence of strong disorder\textemdash for instance, that created
by vacancies, or strongly adsorbed atoms for the same purpose \cite{KPM_G_Ferreira11}.

\begin{figure}[!t]
\centering{}\vspace{0.1in}
 \includegraphics[width=1\columnwidth]{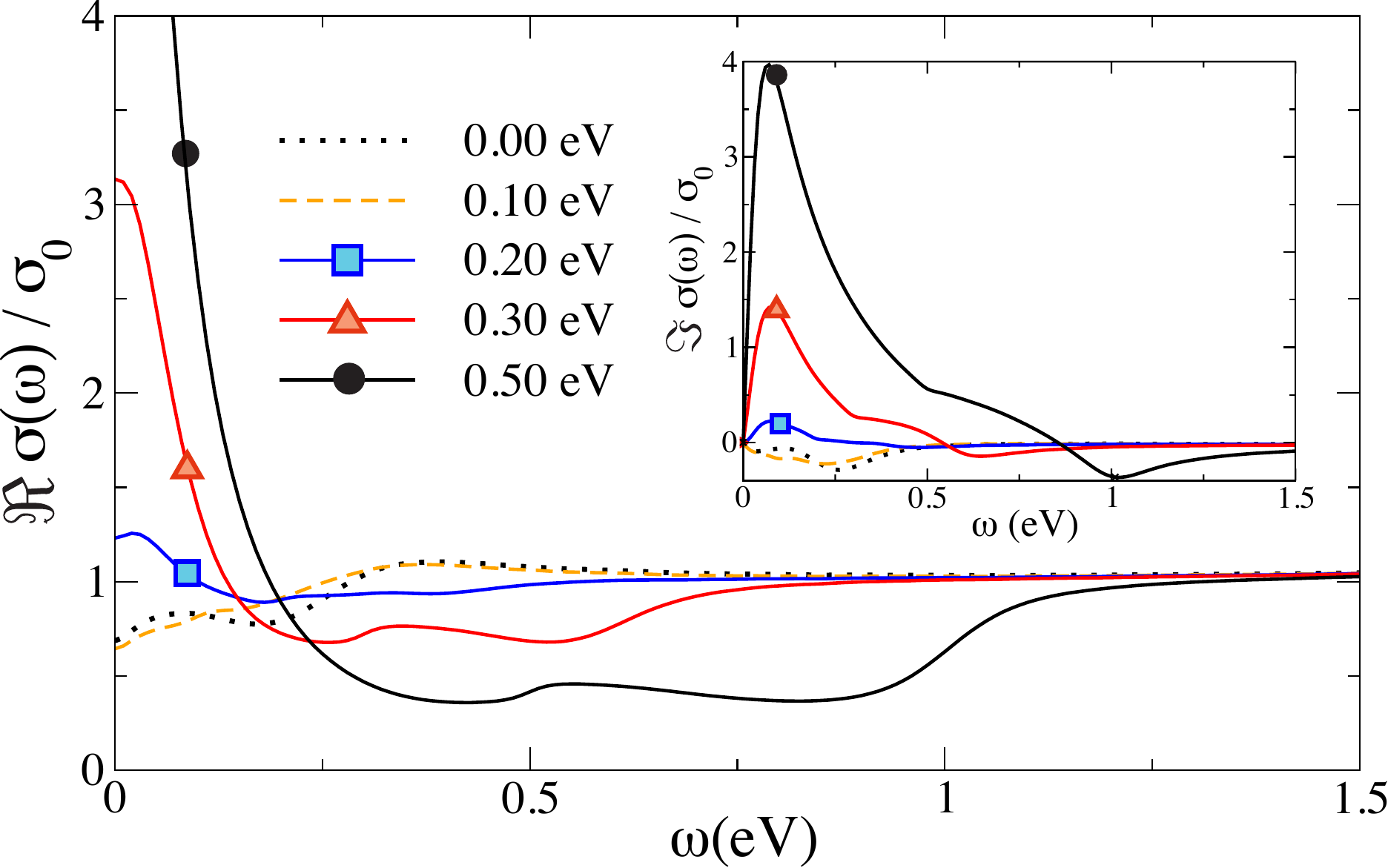} \caption{\label{fig:02_cond}The optical conductivity of graphene with a dilute
vacancy concentration $n_{i}=0.4\%$ at selected values of the chemical
potential $\mu$ with $\eta\approx8$ meV. }
\end{figure}

As a model system of disordered graphene, we have simulated a large
lattice of size $3200\times3200$ (atoms) with a dilute vacancy concentration,
$n_{i}=0.4$\% (atomic ratio). The spectrum of graphene with vacancies
is particle-hole symmetric, and hence for simplicity we assume $\mu\ge0$
in what follows. Owing to the large system size, it suffices to consider
a single disorder realization when performing configurational averages.
The optical conductivity for a typical broadening parameter is shown
in Fig.~\ref{fig:02_cond}. To ensure convergence of the optical
conductivity to a good precision {[}Eq.~(\ref{eq:Re_sigma_Chebyshev}){]},
we have computed a very large number of Chebyshev iterations $N^{2}=8000^{2}$.
Finally, the trace in Eq.~(\ref{eq:Gamma_nm}) has been performed
by means of stochastic trace evaluation (STE) technique \cite{KPM}.
We have used 5000 random vectors in the STE to enable determination
of $\sigma_{nm}$ with accuracy better than $1\%$.

\begin{figure}[t]
\centering{}\vspace{0.1in}
 \includegraphics[width=1\columnwidth]{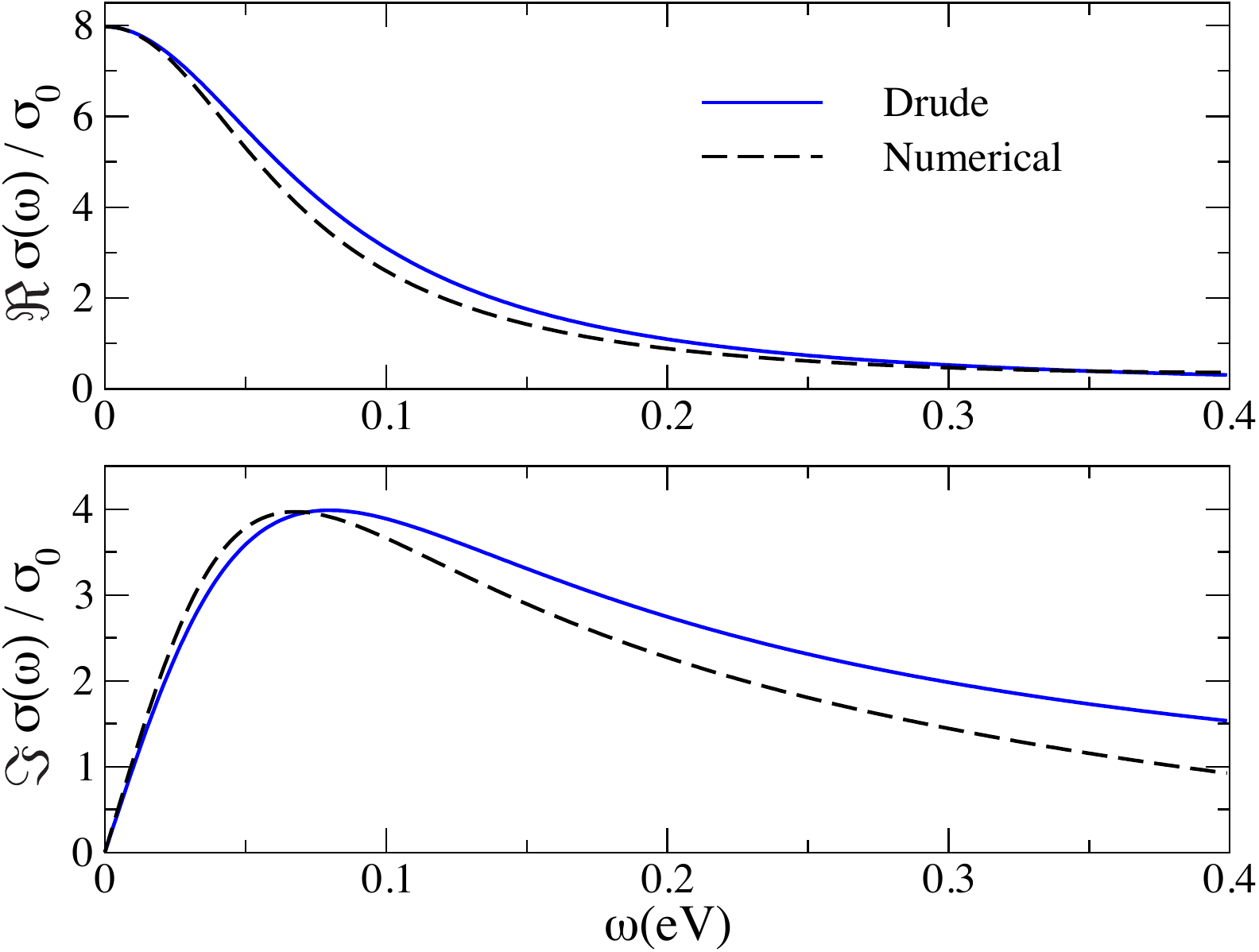}\caption{\label{fig:03_fitDrude}Fit of numerical optical conductivity data with the Drude Model for  $\mu=0.5$ eV$ where \hbar/\tau=0.07$ eV is the adjustable parameter. The spectral weight is given by  $\sigma_{\textrm{D},0}(T,\mu)=\frac{2e^{2}\tau k_{B}T}{\pi\hbar}\log\Big[2\cosh\big[\frac{\mu}{2k_{B}T}\big]\Big]$ }
\end{figure}

Roughly speaking, we expect that disorder should play a role at low
frequencies, $\hbar\omega\ll\mu$. This is the case if the Fermi energy
is not too \textit{small}. Indeed, we see in Fig.~\ref{fig:02_cond}
that for a Fermi energy of 0.5 eV there is a well defined step at
twice the Fermi energy. 
A calculation of the optical conductivity based on the Boltzmann equation,
given by 
\begin{eqnarray}
\sigma_{\textrm{D}}(\omega,T,\mu)=\sigma_{\textrm{D},0}(T,\mu)\frac{1}{1-i\omega\tau},\label{Drude}
\end{eqnarray}
where $\sigma_{\textrm{D},0}(T,\mu)=\frac{2e^{2}\tau k_{B}T}{\pi\hbar}\log\Big[2\cosh\big[\frac{\mu}{2k_{B}T}\big]\Big]$,
predicts the onset of intraband transitions forming a well defined
Drude peak \cite{stauber08,drude}. This becomes clear in Fig.~\ref{fig:03_fitDrude}
where our large-scale numerical calculations for $\mu=0.5$ eV can
be well fit by the Drude model of Eq. (\ref{Drude}) with a single adjustable parameter $\hbar/\tau\approx0.07\ $eV and the spectral weight  $\sigma_{\textrm{D},0}(T,\mu)$.
However, as the Fermi energy decreases, the Fermi step becomes progressively
less well defined (compare, for example, the curves in Fig.~\ref{fig:02_cond}
for 0.3 eV and 0.20 eV; in the latter there is no trace of the Fermi
step). In Eq.~(\ref{Drude}), the intensity of the Drude peak is
proportional, at low temperatures, to $\mu$. Therefore, it is no
surprise that the curves in Fig.~\ref{fig:02_cond} for Fermi energies
of 0.5 eV, 0.3 eV, and 0.2 eV show a progressively smaller intensity
of the Drude peak. Very disordered graphene layers might present a renormalized spectral weight, as observed experimentally in CVD graphene~\cite{CVD}.

However, for smaller Fermi energy the Drude peak is completely washed
out by disorder. In our simulations with a dilute vacancy concentration
(see Fig.~\ref{fig:02_cond}), the critical Fermi energy reads $\mu_{\textrm{c}}\approx0.15$
eV. We note that, in a realistic scenario, the precise value for $\mu_{c}$
will depend on the types and strength of disorder present in the sample. 
The drastic change of behavior in the real part of the optical conductivity
has its counterpart in the imaginary component of this quantity as
ensured by causality. Indeed, for the Fermi energies where the real
part has a well defined Drude peak, one sees in the inset to Fig.~\ref{fig:02_cond}
that the imaginary part of the conductivity changes from negative
to positive as the frequency decreases. This behavior is well known
for the optical conductivity of graphene and signals the dominance
of intraband transitions.

On the contrary, for values of the Fermi energy where the Drude peak
is supressed, the imaginary part of the conductivity is always negative
in the entire frequency range (see Fig.~\ref{fig:02_cond}). This
fact has profound consequences in the interaction of graphene with
electromagnetic radiation. Just to give an example, when $\Im\sigma(\omega)<0$,
graphene does not support $p$-polarized surface waves. On the contrary,
for the case of a well-defined Drude peak both $p-$ and $s-$polarized
waves are supported, albeit in different frequency ranges \cite{ReviewPlasmG_13}.

The reflection coefficients of a graphene sheet are determined by
its optical response. Therefore, we expect that the behavior of the
Casimir\textendash Polder interaction to be strongly dependent on
the details of the optical conductivity, as those discussed above.
Specifically, we expect that for the cases where the Drude peak was
been washed out the curves of the Casimir\textendash Polder interaction
should bunch, whereas for the case where the Drude peak is well defined
such bunching should not occur. This is because, in the former case,
all conductivity curves essentially coalesce among themselves. 
\begin{figure}
\centering{}\vspace{0.1in}
 \includegraphics[width=1\columnwidth]{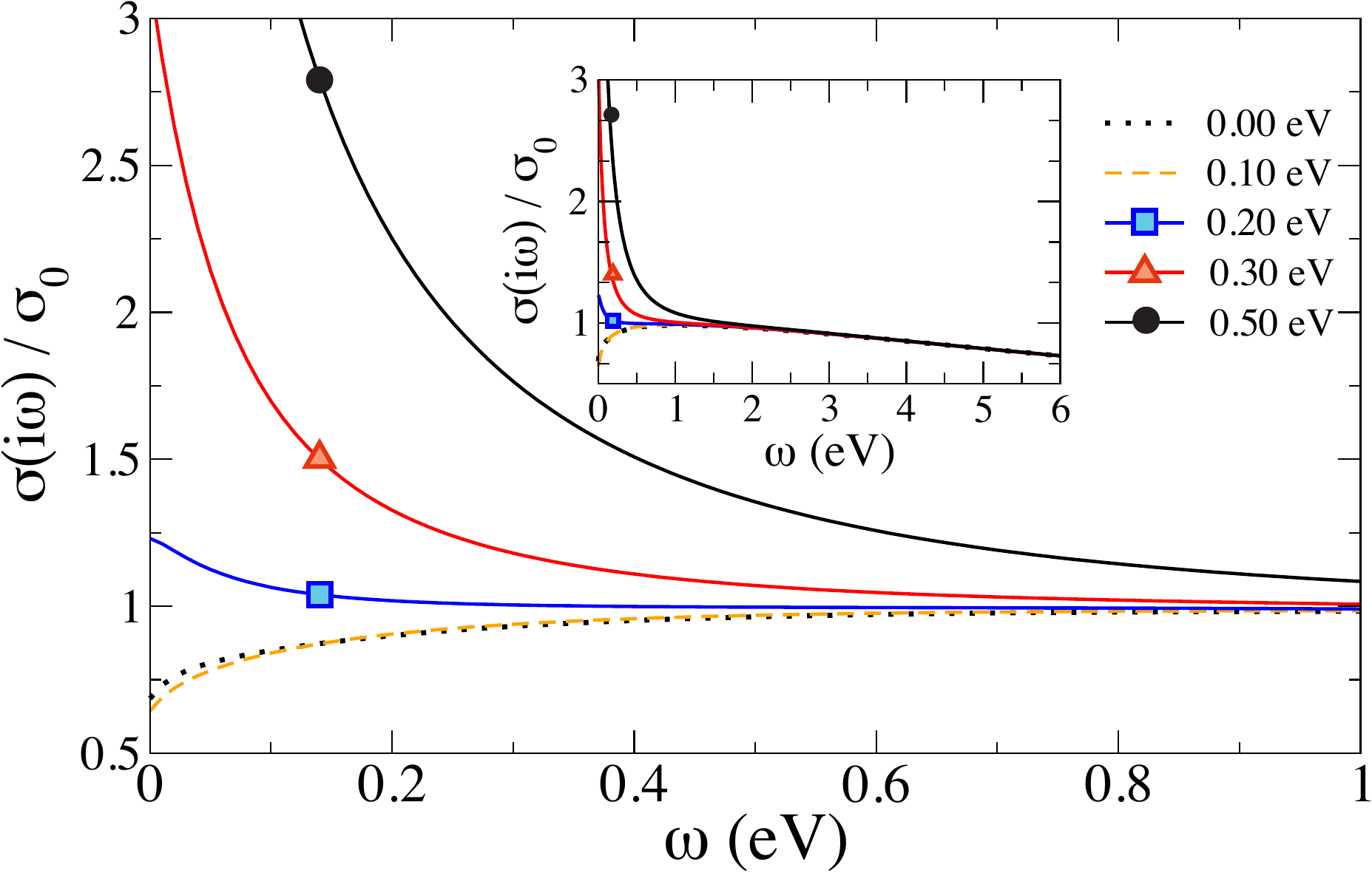} \caption{\label{fig:04_cond_imag_axis}Conductivity in the imaginary frequency
axis at selected values of the Fermi energy. The inset shows the same
curves over a wider range of frequencies.}
\end{figure}

The two regimes discussed above, that is $\mu<\mu^{\ast}$ and $\mu>\mu^{\ast}$,
become quite clear when the optical conductivity is represented in
terms of Matsubara frequencies, as shown in Fig. \ref{fig:04_cond_imag_axis}.
In this figure, the regime where a Drude peak is well define is characterized
by an optical conductivity that presents a positive curvature, whereas
in the opposite case the curvature is negative. Therefore, this way
of representing the optical conductivity data is an effective tool
for separating the two regimes.

\section{Computation of Casimir\textendash Polder interaction\label{sec3}}

Here we compute the Casimir\textendash Polder (CP) energy between
an atom and a graphene sheet with vacancies and discuss the changes
in the CP energy with doping. The optical properties of graphene,
necessary for the calculations, can be well described by the numerical
results presented in the previous section. We consider a rubidium
atom placed at a distance $z$ above a suspended graphene sheet with
chemical potential $\mu$. The whole system is assumed to be in thermal
equilibrium at sufficiently low temperature $T$, such that one can
use the conductivity numerical calculations carried out at $T=0$
K. We choose the rubidium atom due to existence of experimental data
of its electric polarizability for wide range of frequencies \cite{Rubidium}.
The CP energy interaction is calculated within the scattering approach
\cite{Paulo1} 
\begin{eqnarray}
U_{T}(z) & = & \frac{k_{B}T}{\varepsilon_{0}c^{2}}{\sum_{l=0}^{\infty}}'\xi_{l}^{2}\,\alpha(i\xi_{l})\int\frac{d^{2}\textbf{k}}{(2\pi)^{2}}\frac{e^{-2\kappa_{l}z}}{2\kappa_{l}}\nonumber \\
 &  & \hspace{-40pt}\times\Bigg[r^{s,s}(\textbf{k},i\xi_{l},\mu)-\Bigg(1+\frac{2c^{2}k^{2}}{\xi_{l}^{2}}\Bigg)r^{p,p}(\textbf{k},i\xi_{l},\mu)\Bigg]\,,\label{eq:uT}
\end{eqnarray}
where $\xi_{l}=2\pi lk_{B}T/\hbar$ are bosonic Matsubara frequencies,
$\kappa_{l}=\sqrt{\xi_{l}^{2}/c^{2}+k^{2}}$, $\alpha(i\xi)$ is the
electric polarizability of rubidium, and $r^{s,s}(\textbf{k},i\xi,\mu)$,
$r^{p,p}(\textbf{k},i\xi,\mu)$ are the diagonal reflection coefficients
associated with graphene. In Eq.~(\ref{eq:uT}), the prime indicates
that the first term of the summation ($l=0$) is halved.

By modeling graphene as a two-dimensional material with a surface
density current ${\bf K}=\mbox{{\mathversion{bold}\ensuremath{{\sigma}}}}\cdot{\bf E}|_{z=0}$,
and applying the appropriate boundary conditions to the electromagnetic
field, the reflections coefficients are calculated as 
\begin{eqnarray}
 &  & \hspace{0pt}r^{s,s}({\bf k},i\xi,\mu)=\dfrac{2\sigma_{xx}(i\xi,\mu)Z^{h}+\eta_{0}^{2}\sigma_{xx}(i\xi,\mu)^{2}}{-\Delta({\bf k},i\xi,\mu)}\,,\\
 &  & \hspace{0pt}r^{p,p}({\bf k},i\xi,\mu)=\dfrac{2\sigma_{xx}(i\xi,\mu)Z^{e}+\eta_{0}^{2}\sigma_{xx}(i\xi,\mu)^{2}}{\Delta({\bf k},i\xi,\mu)}\,,\\
 &  & \Delta({\bf k},i\xi,\mu)=[2+Z^{h}\sigma_{xx}(i\xi,\mu)][2+Z^{e}\sigma_{xx}(i\xi,\mu)],\label{RefCoefs}
\end{eqnarray}
where $Z^{h}=\xi\mu_{0}/\kappa$, $Z^{e}=\kappa/(\xi\epsilon_{0})$,
$\eta_{0}^{2}=\mu_{0}/\epsilon_{0}$, and $\sigma_{xx}(i\xi,\mu)$
is the longitudinal optical conductivity of graphene \cite{MacdonaldPRL,booknuno}.
In the absence of an external magnetic field, the transverse optical
conductivity of graphene with vacancies vanishes.

A key point for the computation of the CP energy is the correct modelling
of the material surface. In our approach, the characteristics of the
material are incorporated in the longitudinal optical conductivity
$\sigma_{xx}(\omega)$. Far from the Dirac point, Drude's model is
expected to work for frequencies smaller than the chemical potential.
However, for low values of $\mu$, a more accurate calculation must
be carried out to capture the detailed physics of graphene and the
effects of disorder. In our approach, we use the optical conductivity
calculated numerically from a tight-binding Hamiltonian of graphene
with vacancies. For that purpose, we first use the Kramers-Kronig
relations to obtain the conductivities in the imaginary frequency
axis. As shown in Fig.~\ref{fig:04_cond_imag_axis}, in that case,
the separation between the two regimes becomes more clear with different
characteristic curve inflections for each regime at for low frequencies.

\begin{figure}
\vspace{0.1in}
 \centering \includegraphics[width=1\columnwidth]{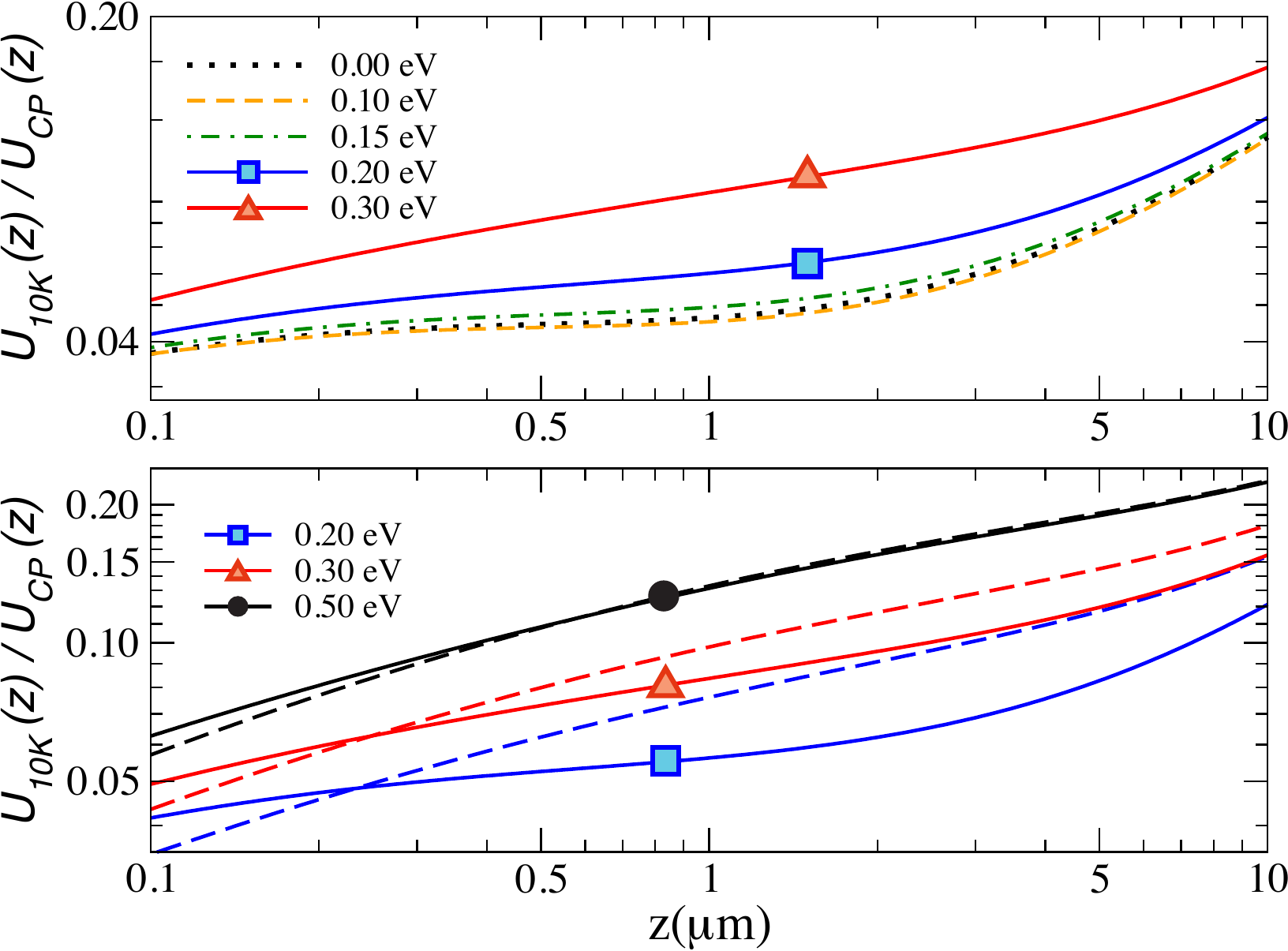} \caption{\label{fig:05_energyCP}Casimir Polder energy between a rubidium atom
and a graphene sheet with $0.4\%$ of randomly distributed vacancies
normalized by the CP energy between an atom and a metal plate as function
of the distance $z$ between the atom and the graphene layer for different
values of the Fermi energy. The lower panel shows the comparison between
the energy obtained with the numerical calculation (solid lines) and
Drude's model (dashed Lines).}
\end{figure}

\begin{figure}[t]
\vspace{0.1in}
 \centering \includegraphics[width=1\columnwidth]{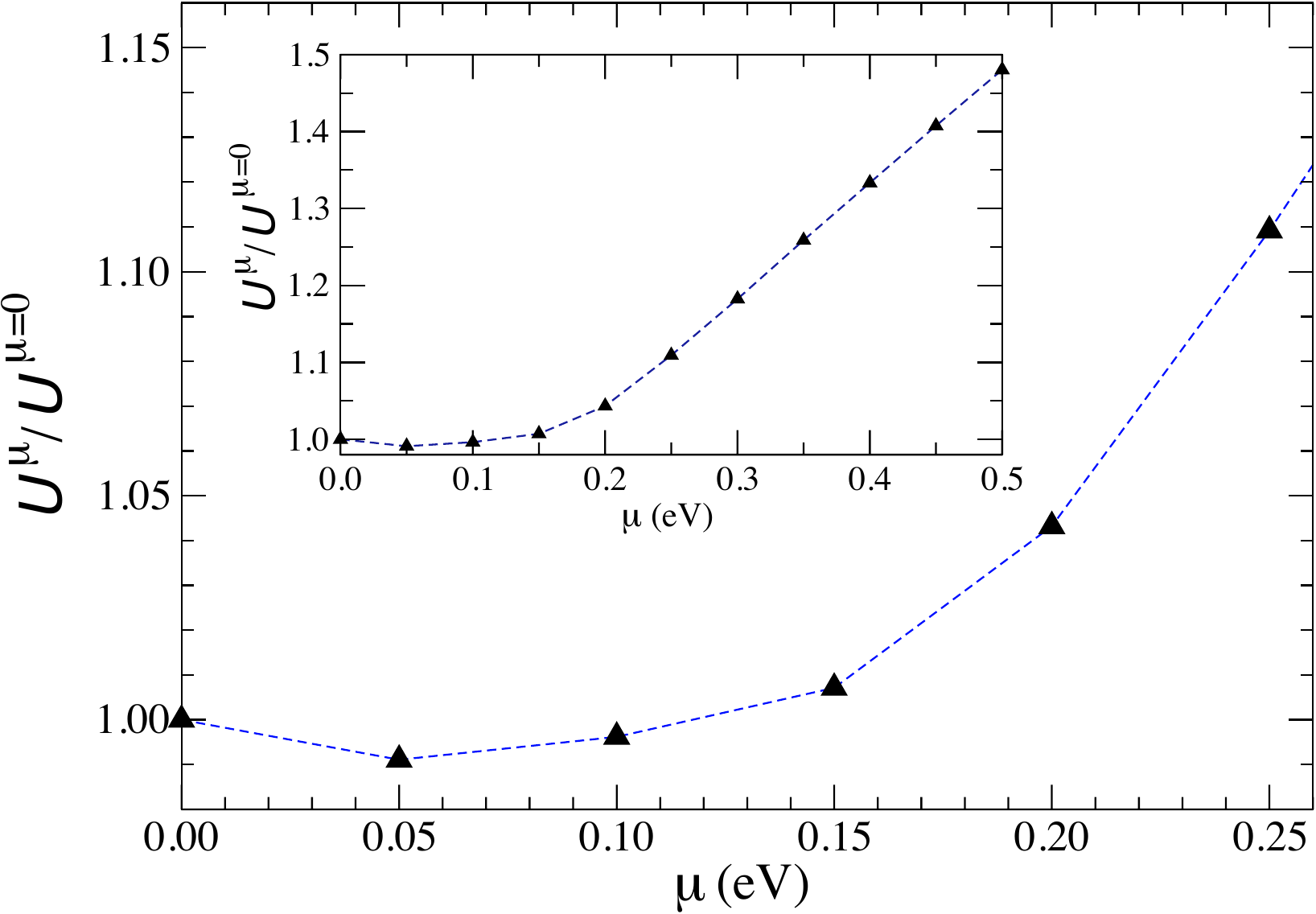} \caption{\label{fig:06_CP}CP energy for finite $\mu$ normalized by CP energy
for $\mu=0$ at $z=2\mu m$ as function of chemical potential.}
\end{figure}

Using equations (\ref{eq:uT}-\ref{RefCoefs}) and the optical conductivities
shown in Fig.~\ref{fig:04_cond_imag_axis}, we calculate the CP interaction
energy between the graphene sheet and a rubidium atom. Figure \ref{fig:05_energyCP}
presents the CP energy normalized by the interaction between an atom
and a perfect metallic surface ($U_{CP}(d)=\frac{-3\hbar c\alpha(0)}{32\pi^{2}\epsilon_{0}d^{4}}$),
as a function of the distance at selected values of $\mu$ and $T$=10K.
For $\mu<\mu^{*}$, graphene behaves basically as a dieletric and
there is a bunching of the CP energy curves. In the opposite regime,
the curves are well separated, as expected from the simple Drude model
{[}Eq.~(\ref{Drude}){]}. The lower panel shows a comparison between
the CP energy calculated using the numerical data and the Drude model
for graphene. Although the optical conductivity curves present a Drude
peak for $\mu>\mu^{*}$, the Drude model does not fit well the numerical
results for the CP energy, overestimating the CP force for experimentally
accessible distances. Saying it differently, our calculations put
stringent constrains on the values of the Fermi energy needed to observe
the CP effect on graphene. If these are too small the force is also
small and one may not be able to measure the effect.

We show in Fig.~\ref{fig:06_CP} the variation of CP energy as function
of the Fermi energy for $z=2\mu$m. The strong effect of the vacancies,
reliably captured by the numerical calculation, results in an almost
constant CP energy for a large range of $\mu$ around the Dirac point.
For larger values of $\mu$, graphene behaves as a Dirac metal, leading
to the linear increase of CP energy as a function of $\mu$ (see inset).
For $\mu$=0.50 eV, the CP energy is increased by $50\%$ if compared
to the neutrality point.

It is clear from our results that the dependency of CP force with
the Fermi energy can be tailored by considering different types of
disorder like adatoms and clusters or a higher concentration of vacancies
and can become a route to manipulate the behavior of dispersive interactions.

\section{Conclusion\label{sec4}}

In this work we have performed realistic large-scale calculations
of the optical conductivity of graphene, revealing the role of disorder
for ``small\char`\"{} Fermi energies. Our calculations show that
in the latter regime, the Drude peak is washed out by disorder and
the application of the Drude conductivity for describing the intra-band
optical conductivity of graphene becomes unjustifiable. This is an
important result, as experiments have been conducted with Fermi energies
around $0.2$ eV, where our calculations show that the Drude model
is no longer valid. As expected, this behavior has important consequences
on the Casimir\textendash Polder effect. For large Fermi energies\textemdash $\mu\sim0.5$
eV\textemdash the optical conductivity of graphene is well described
by a Drude model at low frequencies. However, at ``small\char`\"{}
Fermi energies Drude model breaks down and one cannot distinguished
the Casimir\textendash Polder interaction energies for varying Fermi
energies. Furthermore, the Drude model predicts a larger shift of
the interaction energy relative to that of a perfect metallic plane
that what will happen in a real situation. Therefore, the forces experienced
by the atom will be necessarily smaller than that predicted by the
idealized Drude model and may become difficult to measure. Thus for
a meaningful experiment our study reveals that highly doped graphene
is required.

\section*{Acknowledgements}

T.G.R. acknowledges support from the Newton Fund and the Royal Society
(U.K.) through the Newton Advanced Fellowship scheme (ref. NA150043).
A.F. gratefully acknowledges the financial support of the Royal Society
(U.K.) through a Royal Society University Research Fellowship. N.
M. R. Peres acknowledges the hospitality of UFRJ where this work has
started and financial support from the European Commission through
the project ``Graphene-Driven Revolutions in ICT and Beyond\char`\"{}
(Ref. No. 696656).  T. G. R, N. M. R. P, J. M. V. P. L. thank Brazil Science without Borders program and CNPq for financial support.

\section*{Appendix \label{sec:Appendix.}}

The resolvent operator admits an exact representation in terms of
Chebyshev polynomials \cite{Ferreira15} 
\begin{equation}
(z-\hat{h})^{-1}=\sum_{n=0}^{\infty}a_{n}(z)\,\mathcal{T}_{n}(\hat{h})\,,\label{eq:ident}
\end{equation}
where $z$ is a complex energy variable with $\Im\,z>0$, $\hat{h}$
is a compact Hamiltonian operator satisfying $||\hat{h}||\le1$, and
$\mathcal{T}_{n}(\hat{h})$ are Chebyshev polynomials of first kind
defined by the recursion relations: $\mathcal{T}_{0}(\hat{h})=1\:,\mathcal{T}_{1}(\hat{h})=\hat{h}$,
and 
\begin{equation}
\mathcal{T}_{n+1}(\hat{h})=2\hat{h}\,\mathcal{T}_{n}(\hat{h})-\mathcal{T}_{n-1}(\hat{h})\,.\label{eq:Chebyshev_Matrix_Recursion}
\end{equation}
The expansion coefficients are given by 
\begin{equation}
a_{n}(z)=\frac{2i^{-1}}{1+\delta_{n,0}}\frac{\left(z-i\sqrt{1-z^{2}}\right)^{n}}{\sqrt{1-z^{2}}}\,.\label{eq:exact_coeff}
\end{equation}
These results allow us to express the spectral operator in the form
given in main text {[}Eq.~(\ref{eq:spectral_op_expansion}){]}.


\begin{thebibliography}{10}
\bibitem{reviews} M. Bordag, U. Mohideen, and V.M. Mostepanenko,
Phys. Rep. \textbf{353}, 1 (2001); K.A. Mil-ton, J. Phys. A \textbf{24},
R209 (2004); S.K. Lamoreaux, Rep. Prog. Phys. \textbf{68}, 201 (2005).

\bibitem{casimir} H. B. G. Casimir, Proc. K. Ned. Akad. Wet. \textbf{51},
793 (1948).

\bibitem{casimirpolder} H. B. G. Casimir and D. Polder, Nature (London)
158, 787 (1946).

\bibitem{nem} X. M. H. Huang, C. A. Zorman, M. Mehregany, and M.
L. Roukes, Nature \textbf{421}, 496 (2003).

\bibitem{nem2} M. Aspelmeyer, T. J. Kippenberg, and F. Marquardt,
Rev. Mod. Phys. \textbf{86}, 13911452 (2014). 

\bibitem{LiliaWoods-Review-2015} L. M. Woods, D. A. R. Dalvit, A.
Tkatchenko, P. Rodriguez-Lopez, A. W. Rodriguez, and R. Podgornik,
(2015), arXiv:1509.03338 {[}cond-mat.mtrl-sci{]}.

\bibitem{Galina-Review_2009} G. L. Klimchitskaya, U. Mohideen, V.
M. Mostepanenko, Rev. Mod. Phys. \textbf{81}, 1827 (2009)

\bibitem{rgraphene1} A. H. Castro Neto, F. Guinea, N. M. R. Peres,
K. S. Novoselov, and A. K. Geim, Rev. Mod. Phys. \textbf{81}, 109
(2009).

\bibitem{rgraphene2} N. M. R. Peres, Rev. Mod. Phys. \textbf{82},
2673 (2010).

\bibitem{rgraphene3} K. S. Novoselov, V. I. Falko, L. Colombo, P.
R. Gellert, M. G. Schwab, K. Kim, Nature \textbf{490}, 192 (2012).

\bibitem{Bordag2009} M. Bordag, I.V. Fialkovsky, D.M. Gitman and
D.V. Vassilevich, Phys. Rev. B \textbf{80}, 245406 (2009).

\bibitem{Gomez}G. G\'omez-Santos, Phys. Rev. B \textbf{80}, 245424
(2009).

\bibitem{Svetovoy} V. Svetovoy, Z. Moktadir, M. Elwenspoek and H.
Mizuta, Europhys. Lett. \textbf{96}, 14006 (2011).

\bibitem{Fialkovsky} I.V. Fialkovsky, V.N. Marachevsky and D.V. Vassilevich,
Phys. Rev. B \textbf{84}, 035446 (2011).

\bibitem{Drosdoff} D. Drosdoff and L.M. Wooks, Phys. Rev. A \textbf{84},
062501 (2011).

\bibitem{Sernelius} B.E. Sernelius, Phys. Rev. B \textbf{85}, 195427
(2012).

\bibitem{Bordag2012} M. Bordag, G.L. Klimchitskaya and V.M. Mostepanenko,
Phys. Rev. B \textbf{86}, 165429 (2012).

\bibitem{Galina2013} G.L. Klimchitskaya and V.M. Mostepanenko, Phys.
Rev. B \textbf{87}, 075439 (2013).

\bibitem{Klimchitskaya-Mostepanenko-14} G. L. Klimchitskaya and V.
M. Mostepanenko, Phys. Rev. B \textbf{89}, 035407 (2014).

\bibitem{MacdonaldPRL} W.-K. Tse and A. H. MacDonald, Phys. Rev.
Lett. \textbf{109}, 236806 (2012).

\bibitem{Bordag2016} M. Bordag, I. Fialkovskiy and D. Vassilevich,
Phys. Rev. B \textbf{93}, 075414 (2016)

\bibitem{Drosdoff2012-2} D. Drosdoff, A. D.Phan, L. M. Woods, I.V.Bondarev,
and J.F. Dobson, Eur. Phys. J. B \textbf{85}, 365 (2012).

\bibitem{Galina2014} G. L. Klimchitskaya, U. Mohideen, and V. M.
Mostepanenko, Phys. Rev. B \textbf{89}, 115419 (2014).

\bibitem{Churkin} Yu V. Churkin, A. B. Fedortsov, G. L. Klimchitskaya
and V. A. Yurova, Phys. Rev. B \textbf{82}, 165433 (2010).

\bibitem{Judd} T. E. Judd, R.G. Scott, A.M. Martin, B. Kaczmarek
and T.M. Frohold, New J. Phys.\textbf{13}, 083020 (2011).

\bibitem{Chaichia3} M. Chaichian, G. L. Klimchitskaya, V. M. Mostepanenko
and A. Tureanu, Phys. Rev. A \textbf{86}, 012515 (2012).

\bibitem{Sofia-Scheel-13} S. Ribeiro and S. Scheel, Phys. Rev. A
\textbf{88}, 042519 (2013); Erratum: Phys. Rev. A \textbf{89}, 039904
(2014).

\bibitem{Cysne_2014} T. Cysne, W. J. M. Kort-Kamp, D. Oliver, F.
A. Pinheiro, F. S. S. Rosa , and C. Farina, Phys. Rev. A. \textbf{90},
052511 (2014)



\bibitem{LiliaWoods-2016} N. Khusnutdinov, R. Kashapov, and L. M.
Woods, (2016), arXiv:1602.08443v1 {[}cond-mat.mes-hall{]}


\bibitem{abinitio} A. Ambrosetti, N. Ferri, R. A. DiStasio Jr., and
A. Tkatchenko, Science \textbf{351}, 1171 (2016).

\bibitem{CasimirChern} P. R.-Lopez, and A. G. Grushin, Phys. Rev.
Lett. \textbf{112}, 056804 (2014). 

\bibitem{Huttman-2015} F. Huttmann, \emph{et al}. Phys. Rev. Lett.
\textbf{115}, 236101 (2015)

\bibitem{Ferreira15}A. Ferreira, and E. Mucciolo, Phys. Rev. Lett.
\textbf{115}, 106601 (2015).



\bibitem{RMPgraphene} A. H. Castro Neto, F. Guinea, N. M. R. Peres,
K. S. Novoselov, and A. K. Geim, Rev. Mod. Phys. \textbf{81}, 109
(2009).

\bibitem{Mahan}G. D. Mahan, Many Particle Physics (Plenum Press,
New York, 1981).

\bibitem{Thouless_81}D. J. Thouless and S. Kirkpatrick, J. Phys.
C \textbf{14}, 235 (1981).

\bibitem{Imry}Y. Imry, Introduction to mesoscopic physics, 2nd edition
(Oxford University Press, New York, 2002).

\bibitem{spectral_Kosloff84}H. Tal-Ezer and R. Kosloff, J. Chem.
Phys. \textbf{81}, 3967 (1984).

\bibitem{spectral_Roche_Mayou_97}S. Roche, and D. Mayou, Phys. Rev.
Lett. \textbf{79}, 2518 (1997).

\bibitem{spectral_Tanaka98}H. Tanaka, Phys. Rev. B \textbf{57}, 2168
(1998).

\bibitem{spectral_Weisse04}A. Weisse, Eur. Phys. J. B \textbf{40},
125 (2004).

\bibitem{spectral_Yuan10}S. Yuan, H. De Raedt, and M. I. Katsnelson,
Phys. Rev. B \textbf{82}, 115448 (2010).

\bibitem{spectral_Holzner11}A. Holzner\emph{ et al}., Phys. Rev.
B \textbf{83}, 195115 (2011).

\bibitem{KPM} A. Weisse, G. Wellein, A. Alvermann, and H. Fehske,
Rev. Mod. Phys. \textbf{78}, 275 (2006).

\bibitem{ReviewChebG}N. Leconte, A. Ferreira, and J. Jung, 2D Materials,
Elsevier, Vol.\textbf{ 95}, 35 (2016).

\bibitem{KPM_G_Ferreira11}A. Ferreira \emph{et al}., Phys. Rev. B
\textbf{83}, 165402 (2011).

\bibitem{KPM_G_Fan14} Z. Fan, A. Uppstu, and A. Harju, Phys. Rev.
B \textbf{89}, 245422 (2014)

\bibitem{KPM_G_Garcia15}J. H. Garc\'ia, L. Covaci, and T. G. Rappoport,
Phys. Rev. Lett. \textbf{114}, 116602 (2015).

\bibitem{stauber08} T. Stauber, N. M. R. Peres, and A. K. Geim Phys.
Rev. B \textbf{78}, 085432 (2008). 


\bibitem{drude} N. M. R. Peres, J. M. B. Lopes dos Santos, and T.
Stauber, Phys. Rev. B \textbf{76}, 073412 (2007).

\bibitem{CVD} Jason Horng, Chi-Fan Chen, Baisong Geng, Caglar Girit, Yuanbo Zhang, Zhao Hao, Hans A. Bechtel, Michael Martin, Alex Zettl, Michael F. Crommie, Y. Ron Shen, and Feng Wang
Phys. Rev. B \textbf{83}, 165113 (2011).


\bibitem{ReviewPlasmG_13}Y. V. Bludov, A Ferreira, N. M. R Peres,
M. I. Vasilevskiy, Int. J. Mod. Phys. B \textbf{27}, 1341001 (2013). 

\bibitem{Rubidium} A. Derevianko, S. G. Porsev, and J. F. Babb, At.
Data Nucl. Data Tables \textbf{96}, 323 (2010).

\bibitem{Paulo1} A. M. Contreras-Reyes, R. Gu�rout, P. A. Maia Neto,
D. A. R. Dalvit, A. Lambrecht, and S. Reynaud, Phys. Rev. A \textbf{82},
052517 (2010).


\bibitem{booknuno} P.A.D. Gon\c{c}alves and N. M. R. Peres, \textit{An
Introduction to Graphene Plasmonics}, (World Scientific, 2016, Singapore).




\end{thebibliography}
\end{document}